\documentclass[aps,prb,endfloats,preprint]{revtex4-1}
\usepackage{graphicx,float,amssymb}
\usepackage{dcolumn}
\begin{document}
\title{Electronic structure and magnetism in M(=IA, IIA)C compounds with the rocksalt structure}
\author{Wenxu Zhang}\email{xwzhang@uestc.edu.cn}
\author{Zhida Song, Bin Peng, Wanli Zhang}
\affiliation{SKLETFID, University of
Electronic Science and Technology of China, Chengdu 610054, China}
\begin{abstract}
Electronic structures of MC where M is the alkali and alkaline earth metals with the rocksalt structure are calculated by full potential density functional codes. We find that the spin magnetic moment in the compounds is mainly contributed by the spin polarized p-orbitals of carbon. The large distance between the carbon makes the p-orbitals localized, which induces magnetic instability according to the Stoner criterion. The electronic structure can be mapped to the rigid band model. SrC, BaC and RaC are half-metals, while KC, RbC and CsC are magnetic semiconductors.
\end{abstract}
\maketitle
\section{Introduction}
Spintronics \cite{stsci,strmp} is a very promising subject which can make full use of the spin as well as the electronic properties of electrons where information storage and processing can be performed in the same devices. One of the most crucial materials for spintronics is half-metal, where the electrons at the Fermi level are fully spin polarized. This property allows the spin to be effectively injected and detected. Half-metallicity is theoretically predicted in (half-)Heusler compounds\cite{heusler}, diluted magnetic semiconductor\cite{dms-dp}, oxides and double perovskites\cite{dms-dp,dp}. In order to have this particular property, materials are required to have proper magnetic exchange splitting of the electronic state with different spins to push the Fermi level into the gap of one spin channel. The gap may be developed because of crystal field or fully occupation of certain atomic orbitals. So most known half metals are either oxides, sulfides or Heusler alloys. Magnetism is commonly found in materials with localized d or f electrons. The p-orbitals are generally believed to be delocalized and magnetism is usually not expected. However, recently magnetism is predicted in compounds where p-orbitals are found to host it\cite{volnianska}. The main argument for this p-orbital magnetism is the localization of p-orbitals due to the large interatomic distance. The p-orbital magnetism was intensively investigated in the compounds of (I,II)A-(IV,V,VI)A. Half-metallic ferromagnetism is predicted in compounds of CaC, SrC and BaC with zinc-blende structure\cite{Gao07prb}, as well as with rocksalt structures\cite{Gao07}. The investigation is also extended to rocksalt and zinc-blende LiS, NaS, KS,\cite{gao11} and NaN, KN,\cite{Yan}, alkaline earth selenides\cite{Yogeswari} as well as mononitrides\cite{kozdogan}, pnictides\cite{kozdogan1}. In this work, electronic structures of alkaline earth carbides (M$^\mathrm{IIA}$C) are calculated. Compounds with rocksalt structures are investigated as it was shown that this is more stable than those with zincblende structure\cite{Gao07} at $T=0$. The feature of the electronic structure was mapped to p-orbitals where condition to have half metallicity was discussed based on the rigid band model. The calculations were then extended to alkali metal carbides (M$^\mathrm{IA}$C) where ferromagnetic semiconductor was found in three of these compounds. We also proposed that carbides with mixing of IA and IIA group elements might also be fully spin polarized.
\section{Computational method}
The calculations were performed with the
full-potential local-orbital code\cite{koep} in the version
FPLO9.00-33 with the default basis settings. All calculations
were done within the scalar relativistic approximation. The GGA exchange-correlation
functional was chosen to be PBE \cite{gga}
The number of k-points in the the Brillouin zone was set to $32\times32\times32$
in order to get ensure the convergency of the calculated parameters. The number of k-points is sometimes crucial to the energy and magnetic states \cite{zhangcomm} in these compounds. Convergency of the total energy was set to be better than $10^{-8}$ Hartree together with that of the density better than $10^{-6}$ in the internal unit of the codes. FCC unit cell (space group Fm3m) was used to represent the rocksalt structure with Wyckoff positions assigned to metal (M) and C as M(0,0,0) and C(0.5,0.5,0.5), respectively.
\section{Electron orbital population}
\begin{figure}
  \includegraphics[width=0.5\textwidth]{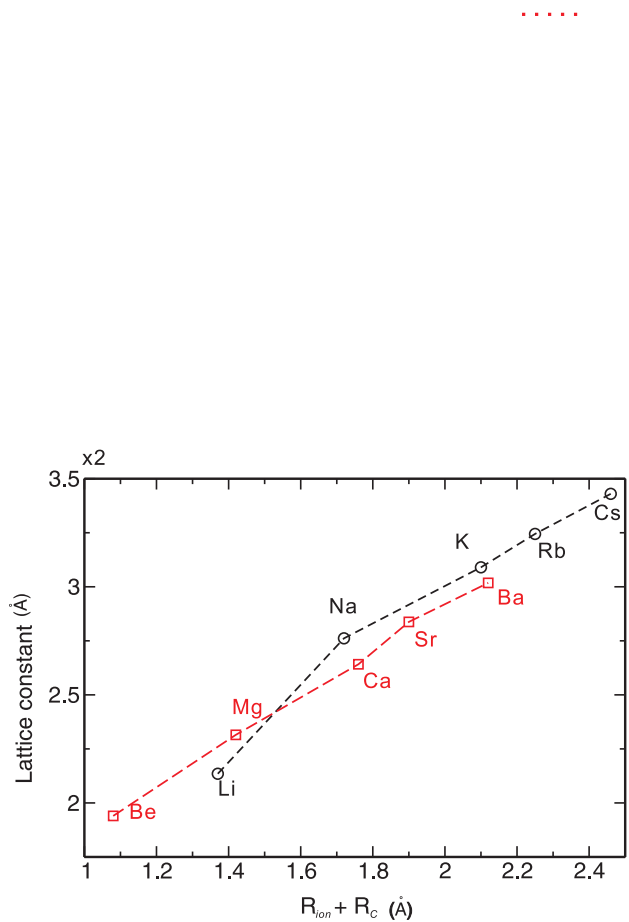}
  \caption{The lattice constant of the compounds at equilibrium plotted as the function of the summation of the ionic radii (R$_{ion}$) of the metallic ions and covalence radius (R$_C$) of C.}\label{fig:lat-radii}
\end{figure}

As shown by the previous works SrC and BaC show half metallic character in the equilibrium ferromagnetic state\cite{Gao07}. The compounds crystalized into the rock-salt structure where the alkaline earth and C atoms form simple cubic lattice, alternately positive and
negative. The lattice constants at equilibrium of the selected M$^\mathrm{IA}$C and M$^\mathrm{IIA}$C compounds are listed in Table \ref{tab:latetc}.
 \begin{table*}
\begin{ruledtabular}
  \caption{Lattice constant at equilibrium ($a_0$), spin magnetic moment ($M$), density of states ($N_0$) at the Fermi level in nonmagnetic states and the exchange integral ($I$) of the compounds and its products with $N_0$ ($IN_0$) }\label{tab:latetc}
  \begin{tabular}{cccccc}
    compounds & $a_0$(\AA)& $M$($\mu_B$)& $N_0$(1/eV)& $I$ (eV)&$IN_0$     \\ \hline
    LiC       &4.27&0   &1.16&0.57&0.66\\
    NaC       &5.52&2.97&2.91&0.54&1.57\\
    KC        &6.18&3.00&6.45&0.53&3.42\\
    RbC       &6.49&3.00&16.09&0.53&8.53\\
    CsC       &6.86&3.00&8.78&0.53&4.65\\

    \hline
    BeC       &3.88&0   &0.84&0.52&0.43\\
    MgC       &4.63&0   &1.78&0.51&0.91\\
    CaC       &5.28&1.92&4.78&0.43&2.06\\
    SrC       &5.68&2.00&4.58&0.42&1.92\\
    BaC       &6.04&2.00&5.40&0.37&2.00\\
    RaC       &6.17&2.00&4.72&0.43&2.03\\
  \end{tabular}
\end{ruledtabular}
\end{table*}
 The lattice constants are roughly proportional to the summation of the ionic radii of the cations and covalent radius of C as shown in Fig.\ref{fig:lat-radii}. This means that the atoms in the compounds can be treated as impenetrable charged spheres. The lattice constant LiC falls off the trend was because of the nonmagnetic nature which favors smaller volume. However, the half of the lattice constant are larger than the summation which means the metal elements are not totally ionic.
 \par  The valance electrons are provided by the 2p orbitals of C and the outmost s electrons from the metals. Projection of the total valence electrons onto each atomic orbitals was shown in Table \ref{tab:electronnum}. The electron population analyse is demonstrating. As the compounds from the same group are similar in electron population, we took BaC and KC as an example for clarity.
\par
\begin{table*}
\begin{ruledtabular}
  \caption{Electron populations projected onto different atomic orbitals of BaC and KC}\label{tab:electronnum}
  \begin{tabular}{ccccc|ccccc}
       &  6s & 6p & 5d & subtotal && 4s & 4p & 3d&subtotal \\
    Ba & 0.24&0.78&0.12& 1.14     &K&0.09&0.08&0.12&0.20\\ \hline
       &  2s &2p&&subtotal       & &2s&2p&&subtotal\\
     C &1.89 &2.97&&4.86       &C&1.91&2.78&&4.69\\
     \hline
     Total&&&&6.00 &&&&&4.97\\
  \end{tabular}
\end{ruledtabular}
\end{table*}
From Table \ref{tab:electronnum}, we see that in BaC the total 6 valence electrons
of Ba 6s$^2$ and C 2s$^2$2p$^2$ are redistributed where 0.86
electron per atom is transferred from Ba to C because of electronegativity and intermixing of the wave functions of the d-orbitals and carbon p-orbitals.. Two 6s electrons of Ba
are partially excited into the 6p and 5d orbitals. Only a small fraction (12\%) is left. This situation is similar in KC. There are totally 0.69 electron transferred from K to C. At the same time one 4s electron is partially excited into the 4p and 3d orbitals. Only 9\% of the s-electrons remains. The 2s electrons of carbon did not participate bonding, which was contributed by the 2p-electrons.
\section{DOS of the compounds}
\begin{figure}
  \includegraphics[width=0.5\textwidth]{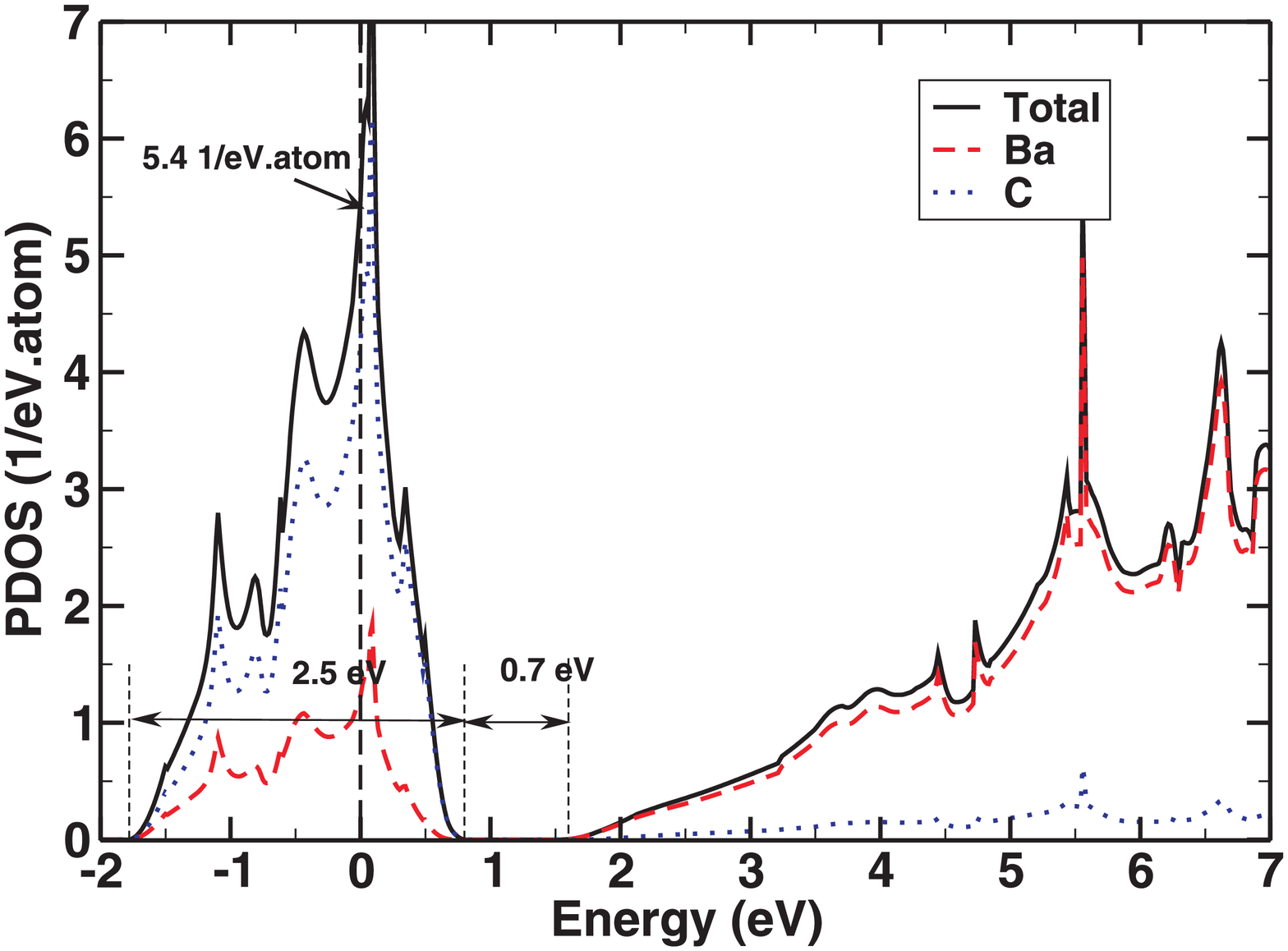}
  \caption{Total DOS in the nonmagnetic state and DOS projected onto different atoms.}\label{fig:dos-nomag}
\end{figure}

\begin{figure}
  \includegraphics[width=0.5\textwidth]{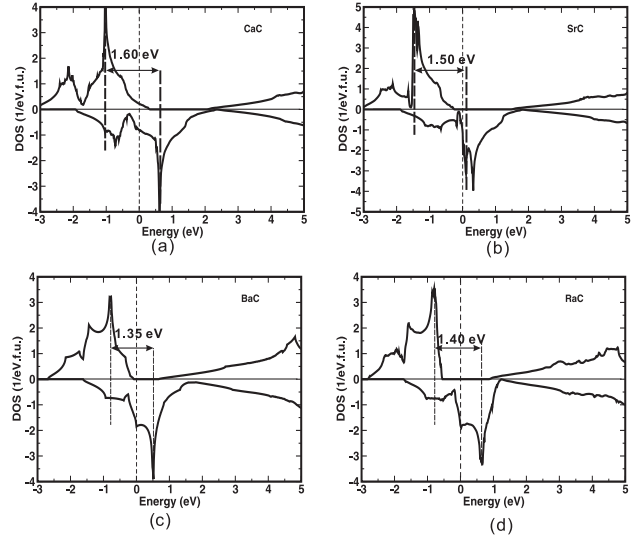}
  \caption{Total DOS of FM states of compounds from IIA metals.}\label{fig:dos-fm-IIAC}
\end{figure}

In order to keep clarity, we begin with the total DOS and its atomic projection in the nonmagnetic state of BaC as shown in Fig. \ref{fig:dos-nomag}. There is a gap of 0.7 eV below and above contributed mainly from C and Ba states respectively. The Fermi level is at the shoulder of the high peak (van Hove singularity) in DOS which is 0.1 eV above the Fermi level. The
DOS at the Fermi level is as high as 5.4 1/eV per formula unit. This high
DOS can lead to magnetic instability according to the Stoner's
criterion as will be discussed later on. Eighty percent of the total DOS at the Fermi level was
contributed by the C atom. The band width is quite narrow, which is
only about 2.5 eV. This is much narrower than the usual bandwidth of the
p-electron, e.g., about 12 eV in diamond. It is due to the much
larger atomic separation of C which is 4.26 \AA\ in BaC while in diamond it is only 1.55 \AA. It can be imaged that when the ferromagnetism sets in, the magnetic exchange splitting pushes the Fermi level in the up spin channel
upwards and that of the down spin channel downwards. Because of the
bandgap, the Fermi level can be set just inside the gap by suitable
exchange splitting. Thus the half metallicity is produced as shown in Fig. \ref{fig:dos-fm-IIAC}(c). The exchange splitting of the up and down spins, estimated from the difference of the maximum and minimum of the spin up and spin down channels, is about 1.35 eV. The other three compounds show similar electronic structure as shown in Fig. \ref{fig:dos-fm-IIAC}. The main difference is the shrink of the bandwidth when going down the row of the Periodic Table. It is because of the increase of the cation radii and naturally the increase of the lattice constant as shown in Fig. \ref{fig:lat-radii}.  The exchange splitting is hardly changed as indicating in the figure. The increase of the bandwidths destroys the half metallicity, and even makes spin polarization unstable as in BeC, where the bandwidth is as large as 9.6 eV (not shown in the figure).
\section{The band structure and the model DOS}
\par
\begin{figure}
  \includegraphics[width=0.5\textwidth,angle=0]{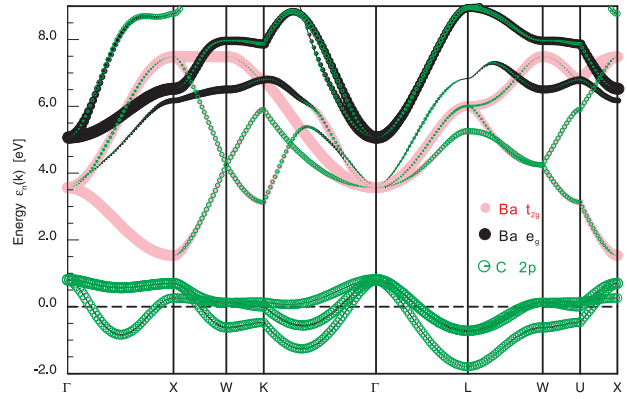}
  \caption{Bands and fat bands of BaC, where the bands from Ba 5d was projected into IR of the site symmetry, namely t$_{2g}$ and e$_g$.}\label{fig:fatbands-nomag}
\end{figure}

 In order to understand the electronic structure more clearly, the so-called fat bands together with the normal bands are shown in Fig. \ref{fig:fatbands-nomag}. It can be seen that the states around the Fermi level are mostly from C 2p orbitals. The van Hove singularity in the DOS in Fig. \ref{fig:dos-nomag} about 0.1 eV above the Fermi level is from the non-dispersing bands along $K-W-U$ directions in the reciprocal space. There are some contributions to the states around the Fermi level from the t$_{2g}$ states of Ba 5d because of unavoidable interactions between Ba and C. The states above the gap are mainly from 5d states of Ba. At the $\Gamma$ point, the $e_g$ and $t_{2g}$ states are separated by 1.5 eV. The bonding states at $X$ from $t_{2g}$ set an upper energy of the gap. The nonmagnetic compound is an indirect bandgap semiconductor.
\par
\begin{figure}
  \includegraphics[width=0.5\textwidth]{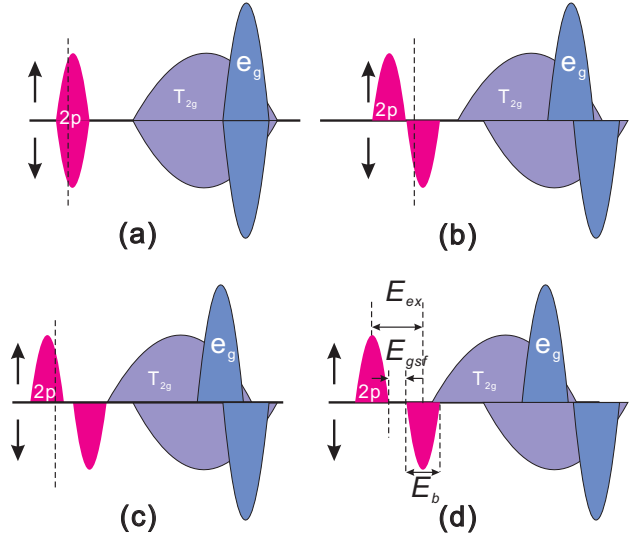}
  \caption{Model DOS of the compounds in nonmagnetic state (a), magnetic state when M = IA (b), magnetic state when IIA elements are substituted by IA elements (c), and magnetic state when M = IIA (d) The arrows pointing up and down represent state of spin up and down respectively. The dashed vertical lines in (a), (b) and (c) indicate the Fermi level.  }\label{fig:model-dos}
\end{figure}

 From the analyse above, we can schematically draw the DOS of the compounds as shown in Fig. \ref{fig:model-dos}. In the nonmagnetic state the 2p states locate around the Fermi level. The d states from the metal are split into e$_g$ and t$_{2g}$ bands. They are separated from the 2p states by an energy gap.  The DOS of spin down and spin up are equally occupied as in Fig. \ref{fig:model-dos}(a). According to this model, six electrons can be hold in 2p orbitals at most. The magnetic moment $M$ (in $\mu_B$) and the total number
of the valence electrons N$_{val}$ are related by
\begin{equation} M=3-(N_{val}-3)=6-N_{val}.
\label{equ:mom-val}
\end{equation}
The total valence electrons include the outmost s-electrons in the metals and two 2p-electrons in C. The 2s electrons of C do not participate bonding because of its localization in these compounds. In this rigid band model, the exchange splitting ($E_{ex}$), the spin flip bandgap $E_{gsf}$ which is defined by the difference of the lowest unoccupied down spin level and the highest occupied up spin level and the bandwidth ($E_{b}$) of the p-orbitals are related by
\begin{equation} E_{ex}=E_{gsf}+E_{b}.
\label{equ:egb}
\end{equation}
According to this model, there are two possibilities to obtain half metallicity: the spin up channel is fully occupied and the spin down channel is partially occupied as in Fig. \ref{fig:model-dos}(b) or the spin up channel is occupied and the spin down channel is empty as in Fig. \ref{fig:model-dos}(c). There is a special case where the spin up channel is full and the spin down is empty, so that magnetic semiconductor with different gaps can be obtained as shown in Fig. \ref{fig:model-dos}(d). The DOS model shown in (b) is the case of SrC, BaC, RaC as shown in Fig. \ref{fig:dos-fm-IIAC}. In the cases of (c) and (d), there should be less electrons in the system. Thus if we substitute IIA elements with IA ones, we can have situation (d) because 3 electrons in the valence state can fully occupy the spin up channel. We can arrive (c) by partial substitution of IIA by IA elements.
\section{DOS of the M$^{IA}$C compounds}
\begin{figure}
  \includegraphics[width=0.5\textwidth]{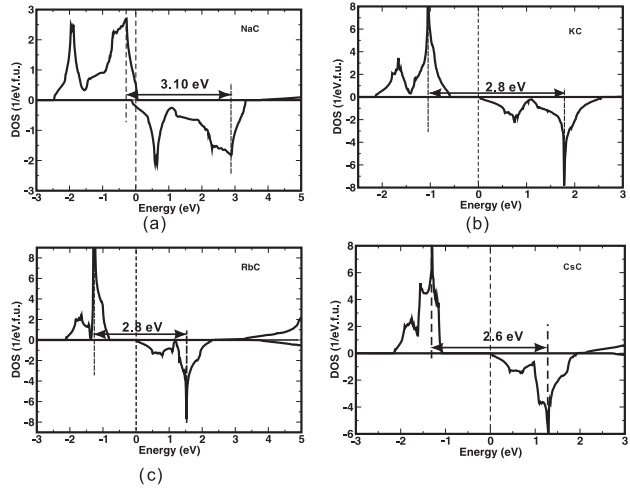}
  \caption{Total DOS of FM states of compounds from M$^\mathrm{IA}$C.}\label{fig:dos-fm-IAC}
\end{figure}

 Motivated by the model DOS analysis above, we performed calculation to M$^\mathrm{IA}$C. The DOS of MC compounds where M = Na, K and Rb are shown in Fig. \ref{fig:dos-fm-IAC}. LiC is nonmagnetic and the DOS is omitted here. From the figure, we see that KC, RbC and CsC are semiconductors with fully spin polarized 2p states, and the spin flipping gap are 0.5 and 0.8 eV, respectively. NaC is on the verge of being fully spin polarized. It is due to the large bandwidth compensating the enhancement of the exchange splitting. The spreading of the p-orbitals is largely depended on the atomic distances, scaling with $d^{-2}$ according to Harrison \cite{harrison} where $d$ is the atomic distance. In this case the exchange splitting is not large enough to push the down spin states to higher energy and transfer the down spin electrons to the up spin channel to make it fully occupied. In NaC, metallic state is induced because the large spreading of the p-state which is more than 3.5 eV due to the small lattice constant of 5.52 \AA\ compared with the other magnetic compounds in this row. C's in LiC are much too close so that the DOS at the Fermi level is only about 1.16 eV$^{-1}$ which is far too low to intrigue magnetic instability as will be discussed below.
\section{Stoner criterion on p-orbital magnetism}
The magnetism in the compounds originates from the relatively
contracted wave functions of the p-orbitals and its unsaturation as
discussed by Volnianska\cite{volnianska,sieberer}. The relative large atomic radii of the cations
reduce the hopping of the p-electrons of the carbon. The bandwidth read from the DOS is 3.0 eV which is at the same scale and even smaller than that of 3d elements. This is an indication of the localization of the orbitals. This narrow
spread of the p-state results in the high DOS at the Fermi level, which
induces magnetic instability according to the Stoner criterion. In the simplified version of this theory the Stoner criterion for spin polarization is $IN_0>1$ where $I$ is the exchange integral (Stoner parameter) and $N_0$ is the DOS at the Fermi level at the ground state. The exchange integral determined from \begin{equation}
\chi=\frac{N_0}{1-IN_0}
\label{equ:exchange}
\end{equation}
where the susceptibility is calculated by
\begin{equation}
\frac{1}{\chi}=\frac{\partial^2 E(m)}{\partial m^2}.
\label{equ:chi}
\end{equation}
The $E(m)$ curves in Equ. (\ref{equ:chi}) are calculated by the fixed spin moment method. The DOS at the Fermi level ($N_0$) and the exchange integrals ($I$) estimated from Equ. (\ref{equ:exchange}) and their products ($IN_0$) of the compounds are listed in Table \ref{tab:latetc}. As shown in the Table, the products ($IN_0$) are larger than 1, as it should be, in the compounds with magnetic ground state. In these two groups of compounds the exchange integral are essentially not changed. However, the exchange splitting in IA group is larger than that in IIA, which is the consequence of larger magnetic moment, because the exchange splitting is proportional to $Im^2$. Larger exchange constants can generally lead to higher Curie temperature according to the mean field consideration, which is favorable in applications.
\section{Conclusions}
In summary we have used full potential density functional codes to study the electronic structure of alkali and alkaline earth carbides with the rocksalt structure. It was found that SrC, BaC, RaC were halfmetals, while KC, RbC and CsC were semiconductors. It was also possible to have half-metallic electronic structure by doping alkali compounds with alkaline earth or vice versa according to our analysis. The spin polarization was mainly within the localized p-orbitals in C due to the large lattice constant determined by the large radii of the cations. The origin of the magnetism could be understood from the Stoner's itinerant theory where the exchange integrals of the compounds were calculated. The electronic structure could be explained by a rigid band model. These compounds are similar to these with the zincblende structure\cite{Gao07prb}: The carbon atoms form fcc sublattice with the lattice constants only about 10\% smaller than those of the zincblende structure at equilibrium. The density of states of both show two humps mainly from the p-states due to the ligand field. The main features determined by the carbon atoms gives an opportunity to tune the magnetic exchange splitting by applying pressure where spin gapless semiconductor may be realized, which may have superior electronic properties. Although these compounds are predicted to show attractive properties for spintronics, they are hypothetical at present. It may be a challenge to synthesize the compound, but it is highly recommended.
\section{Acknowledgement}
This work was financially supported by International Science \& Technology Cooperation Program of China(2012DFA510430) and Excellence of Central Universities.


\end{document}